\documentclass[12pt,preprint]{aastex}

\newcommand{\myemail}{s-a-chin@tamu.edu}
\newcommand{\be}{\begin{equation}}
\newcommand{\ee}{\end{equation}}
\newcommand{\ba}{\begin{eqnarray}}
\newcommand{\ea}{\end{eqnarray}}
\shorttitle{Forward Symplectic Algorithms}
\shortauthors{S. A. Chin and C. R. Chen$^*$}
\begin{document}
\title{Forward Symplectic Integrators for Solving \\Gravitational Few-Body Problems}
\author{Siu A. Chin and C. R. Chen}
\affil{Center for Theoretical Physics, Department of Physics,\\ 
              Texas A\&M University, College Station, TX 77843}
\email{\myemail}
\begin{abstract}
  
We introduce a class of fourth order symplectic algorithms 
that are ideal for doing long time integration of gravitational few-body 
problems. These algorithms have only positive time steps, but require 
computing the force gradient in additional to the force. 
We demonstrate the efficiency of these Forward Symplectic
Integrators by solving the circularly restricted three-body problem in the 
space-fixed frame where the force on the third body is explicitly 
time-dependent. These algorithms can achieve accuracy of Runge-Kutta,
backward time step and corrector symplectic algorithms
at step sizes five to ten times as large.
\end{abstract}

\section{Introduction}

Symplectic Integrators (SI) \citep{yos93,cha96} conserve all 
Poincar\'e invariants when integrating classical trajectories. 
For periodic orbits, their energy errors are bounded and periodic, 
in contrast to Runge-Katta type algorithms whose energy error increases 
linearly with integration time \citep{gla91}. Symplectic algorithms are therefore 
better long time integrators of astrophysical many-body problems 
\citep{wis91,saha92} and have been studied extensively in the literature 
\citep{can91, yos93, mcl95, kos96, bla99}. However, current higher order
SI seem to suffer two disadvantages.
First, the number of force evaluations require by these algorithms
proliferates much more rapidly than non-symplectic algorithms. For example, 
a six order Runge-Kutta-Nystr\"om algorithm \citep{alb55} requires 
only five force evaluations; a six order symplectic algorithm with negative 
intermediate time steps requires at least seven force evaluations \citep{yos90}. 
Second, they seem to have a shorter stability range. This may also due to 
the use of negative time steps for cancelling lower order errors. 
In this work, we consider a class of fourth order Forward Symplectic 
Integrators (FSI), which have only positive time steps \citep{chin97,chin02}. 
(We have referred to these as gradient symplectic algorithms previously.) 
The price for having only positive time steps is that one must 
compute the force gradient in additional to the force. This additional 
effort for gravitational few-body problems is very modest, but the
resulting gain in numerical efficiency and stability is tremendous. 
We will show below that these FSI can achieve accuracy of existing 
symplectic integrators at step sizes five to ten times as large. 
These very powerful algorithms should be considered by colleauges 
doing long time few-body simulations. 

Despite the fact that \citet{wis91} have questioned in 1991
why there were no FSI beyond second order,
this class of algorithms has been unrecognized and overlooked until recently. 
The reason for this is simple; classical dynamics is time reversible, 
there is no compelling demand for purely positive time steps. The demand for 
positive time steps only arises when one has to solve {\it time-irreversible} 
equations, such the diffusion \citep{auer01}, Fokker-Planck \citep{for01}, 
and Navier-Stoke equations. However, once FSI have been derived, they can be used
to solve {\it both} time-reversible and time-irreversible evolution equations. 
For time-irreversible systems, they are the only decomposition algorithm 
possible, but even for time-reversible equations, they give rise to 
algorithms of great efficiency and stability.    
  
In the next Section, we will describe in detail how this class of 
algorithms fits into the large picture of symplectic integrator
development. We will show that our FSI traces it lineage to Ruth's seldom 
used third order algorithm \citep{ruth83}. In Section 3, we describe our fourth order
FSI and the one-parameter family of integrators. In Section 4, we compare various 
algorithms by computing a lengthy and intricate closed orbit of a 
circularly restricted three-body problem. Physical three-body problems, 
such as the Sun-Earth-Moon, or Sun-Jupiter-Saturn, are too ``tame" for 
discriminating the merit of these very powerful algorithms. Here, we 
also note a puzzling fact that some algorithms can appear to converge at 
higher order than expected. We summarize our conclusions and 
suggest some future direction of research in Section 5. 

\section{Relation to Existing Symplectic Integrators}

To give context to our work,
we will review the development of symplectic integrators in the
general contex of solving any evolution equation of the form
\be
{{\partial w}\over{\partial t}}=(T+V)w,
\label{gen}
\ee
where $T$ and $V$ are non-commuting operators.
Classically, the evolution of a dynamical variable $w(q_i,p_i)$ is
given by the Poisson bracket (sum over repeated indices),
\begin{equation}
{{d }\over{dt}}w(q_i,p_i)=\{w,H\}\equiv
         \Bigl(
		          {{\partial w}\over{\partial q_i}}
                  {{\partial H}\over{\partial p_i}}
				 -{{\partial w}\over{\partial p_i}}
                  {{\partial H}\over{\partial q_i}}
				                    \Bigr).
\label{peq}
\end{equation}
For a Hamiltonian of the form, 
\begin{equation}
H(p,q)={1\over{2}}p_ip_i+v(q_i),
\label{ham}
\end{equation}
equation (\ref{peq}) is of the form (\ref{gen})
with $T$ and $V$ given by 
\begin{equation}
T={p_i}{{\partial }\over{\partial q_i}},\qquad
V=F_i{{\partial }\over{\partial p_i}},
\label{tandv}
\end{equation}
where $F_i=-\partial v/\partial q_i$. For this case, 
the constituent evolution operators
${\rm e}^{\epsilon T}$ and ${\rm e}^{\epsilon V}$ 
displace $q_i$ and $p_i$ forward in time
via 
\begin{equation}
q_i\rightarrow q_i+\epsilon p_i\quad 
{\rm and}\quad p_i\rightarrow p_i+\epsilon F_i.
\label{maps}
\end{equation}
In general, the evolution equation (\ref{gen}) can be solved iteratively via
\be
w(t+\epsilon)={\rm e}^{\epsilon(T+V)} w(t),
\label{two}
\ee
provided that one has a suitable approximation for the short 
time evolution operator ${\rm e}^{\epsilon(T+V)}$. If
${\rm e}^{\epsilon T}$ and ${\rm e}^{\epsilon V}$ can be exactly 
implemented individually, then ${\rm e}^{\epsilon(T+V)}$ 
can be decomposed to arbitrarily high order in the form
\be
{\rm e}^{\epsilon (T+V )}=\prod_i
{\rm e}^{a_i\epsilon T}{\rm e}^{b_i\epsilon V}.
\label{prod}
\ee
For example, from the second order factorization
\be
{\rm e}^{\epsilon(T+V)}
\approx
{\rm e}^{{1\over 2}\epsilon T}
{\rm e}^{\epsilon V}
{\rm e}^{{1\over 2}\epsilon T}={\cal T}^{(2)},
\label{second}
\ee
one can construct a fourth order approximation via
\be
{\cal T}_{FR}^{(4)}(\epsilon)={\cal T}^{(2)}(\widetilde\epsilon)
{\cal T}^{(2)}(-s\widetilde\epsilon)
{\cal T}^{(2)}(\widetilde\epsilon),
\label{fr}
\ee
where
$s=2^{1/3}$ is chosen to cancel the third order error in ${\cal T}^{(2)}$, 
and $\widetilde\epsilon=\epsilon/(2-s)$ rescales
the sum of forward-backward-forward time steps back to $\epsilon$.
This fourth order Forest-Ruth (FR) scheme \citep{for90} has been independently 
derived many times \citep{cam90,can91}. The above construction, while widely 
known through the work of \citet{yos90}, 
was first published by \citet{cre89}.  

The middle time step in (\ref{fr}) is negative. This is not
accidental. \citet{she89} and \cite{suz91} have independently proved that, 
beyond second order, any factorization of the form (\ref{prod}) must contain 
some negative coefficients in the set $\{a_i, b_i\}$. \citet{gol96}
later proved that any factorization of the form (\ref{prod}) must contain at least 
one negative coefficients for {\it both} operators. This means that for any
evolution equations containing an irreversible operator, such as the diffusion
kernel $T={1\over{2}}\nabla^2$, no algorithms of the form (\ref{prod}) is possible 
beyond second order. This is because 
$\langle{\bf r}^\prime|{\rm e}^{a_i\epsilon T}|{\bf r}\rangle
\propto {\rm e}^{-({\bf r}^\prime-{\bf r})^2/(2a_i\epsilon)}$ is the 
diffusion kernel only if $a_i$ is positive.
If $a_i$ were negative, then the kernel would be unbound and unnormalizable,
reflecting the fact that diffusion is a time-irreversible process. 
Positive coefficients are therefore absolutely essential for  
solving any evolution equation having an irreversible component. 
We have shown, and will further domonstrate below, that 
even for time-reversible systems such as quantum \citep{chin02} 
or classical dynamics \citep{chin97}, 
purely positive coefficients give rises to very efficient algorithms with
excellent convergence properties.

While Sheng's proof is slightly more general, Suzuki's proof provides insight 
into how to circumvent this negative coefficient problem. The essence of 
Suzuki's proof is to note that, for example, one has the following
operator respresentation of the velocity form of the Verlet algorithm, 
\be
{\rm e}^{{1\over 2}\epsilon V}{\rm e}^{\epsilon T}{\rm e}^{{1\over 2}\epsilon V}
={\rm e}^{\epsilon H^\prime},
\ee
with
\be
H^\prime=T+V-{1\over 12}\epsilon^2[T,[V,T]]
               +{1\over 24}\epsilon^2[V,[T,V]]+O(\epsilon^4). 
\label{velver}
\ee
In order to obtain a fourth order algorithm, one must eliminate the two 
double commutator error terms above. With purely positive coefficients 
$a_i$ and $b_i$, one can eliminate either one,  but not both. 
Thus to obtain a fourth order factorization with only positive coefficients,
one must {\it keep} one of the two double commutators. With 
$T$ and $V$ as defined by (\ref{tandv}), the commutator
\begin{equation}
[V,[T,V]]=2F_j{ {\partial  F_i}\over{\partial q_j} }
                { {\partial  }\over{\partial p_i} }=
                \nabla_i(|{\bf F}|^2)
                { {\partial  }\over{\partial p_i} },
\label{vtv}
\end{equation}
simply give rises to a new force defined by the gradient of the square of the 
force modulus. This commutator can therefore be kept. The use 
of this force gradient is not new. \citet{ruth83} in his pioneering paper 
paper has derived a third order derivative algorithm on the basis of 
canonical transformations. However, this algorithm was universally 
ignored \citep{cha96} with no follow-up discussion. In 1997 one of us  
\citep{chin97} noted that Ruth's algorithm actually corresponds to a 
time-asymmetric operator decomposition and that his force derivative is 
precisely the force gradient given by $[V,[T,V]]$.  
The importance of Suzuki's proof is that it 
draws a clear connection between the demands for positive time steps 
and the necessity of keeping higher order commutators, such as $[V,[T,V]]$. 

If one were to keep $[V,[T,V]]$, then there are two distinct ways of 
eliminating the other double commutator $[T,[V,T]]$.  
Despite the lack of an operator formalism, \citet{row91} 
also noted that the velocity form of 
the Verlet algorithm has error terms of the form (\ref{velver}). 
He proposed to get rid of the $[T,[V,T]$ term by an explicit transformation
and to view the algorithm  
as fourth order for a Hamiltonian with a modified potential 
\be
V^\prime=V+{1\over 24}\epsilon^2[V,[T,V]].
\label{modforce}
\ee
The reinterpretation of the potential is easy, the tricky part here 
is that since one has performed a transformation 
to eliminate $[T,[V,T]]$, one must transform back correctly to preserve 
fourth order accurracy for the original variables. This transformation
is often more complicated than the original algorithm. This way of 
eliminating $[T,[V,T]]$ is tantamount to applying operators
\be
{\rm e}^{C(\epsilon)}
{\rm e}^{{1\over 2}\epsilon V}{\rm e}^{\epsilon T}{\rm e}^{{1\over 2}\epsilon V}
{\rm e}^{-C(\epsilon)}
\ee
where ${\rm e}^{-C(\epsilon)}$ and ${\rm e}^{C(\epsilon)}$ are the initial 
and final transformations respectively. They can be further decomposed
into the product form, 
${\rm e}^{C(\epsilon)}=\prod_i{\rm e}^{c_i\epsilon T}{\rm e}^{d_i\epsilon V}$.
This class of ``process" \citep{mar96,mar97,bla99} or 
``corrector" \citep{wis96,mcl96} symplectic algrithms has the distinct advantage that 
when one iterates the algorithm, only the kernel algorithm needs to be 
iterated. If there is no need to keep track of intermediate results, 
then a fourth order algorithm is possible with 
only a single evaluation of the modified force \citep{mar96,mar97,wis96,bla99}. 
However, by its very construction, either ${\rm e}^{-C(\epsilon)}$ or 
${\rm e}^{C(\epsilon)}$ must contain negative time steps,
and cannot be applied to equations with an irreversible operator.
Nevertheless they are valuable addition to the symplectic repertoire for
solving reversible dynamical problems. More recent 
process algorithms also uses a modified potential, but only in the 
form (\ref{modforce}) and only in the context of a second order kernel 
algorithm \citep{bla99}. \citet{mar97} have used a two-fold Rowland 
kernel that could have been fourth order, but the correct parameter 
values for that were never considered. 

In the operator formalism, a more direct way of eliminating 
$[T,[V,T]]$ is to add more operators symmetrically. In this approach, the
elimination of $[T,[V,T]]$ is built-in, with no need of an extrinsic
transformation. \citet{mcl952} has done this for algorithms 4A and 4B 
(described below) in the context of a slightly perturbed Hamiltonian system. 
But his algorithms were not truly fourth order because 
he did not bother to {\it keep} the commutator $[V,[T,V]]$. These pseudo-higher 
order algorithms \citep{mcl952,cha00,las00}, which only require a single error commutator 
to vanish at each order, are much simpler than regular higher order 
algorithms, which require {\it all} error commutators to vanish at each order.  
There is also a widely cited reference by \citet{kos93} that 
purported to contain a fourth order scheme using modified potentials.
Koseleff's algorithm would have been algorithm 4A discussed below, 
unfortunately, his coefficient for the commutator $[V,[T,V]]$ is 
{\it incorrect}. It should be 1/72 rather than 1/24.  

\section{Fourth Order Forward Symplectic Integrators}

Suzuki, on the basis of McLachan's work \citep{mcl952}, retained the 
commutator $[T,[V,T]]$
and wrote down two fourth order factorization schemes 4A and 4B with purely 
positively coefficients \citep{suz96}. He did not implement them
classically or quantum mechanically, nor in anyway demonstrated their
usefulness. One of us \citep{chin97} derived schemes 4A and 4B by elementary means 
independent of McLachan's work, and have explicitly implemented them 
classically and demonstrated their effectiveness in solving the Kepler problem.
Moreover, algorithms 4C \citep{chin97}, which is the direct 
descendant of Ruth's third order algorithm \citet{ruth83}, in all cases tested, has 
have much smaller error coefficients than 4A and 4B. Prior to Chin's work, 
we are not aware of any implementation of fourth order FSI for solving any problem.

Since the commutator and the gradient force term occur frequently in the
following, we will define
\be
U(t)\equiv[V(t),[T,V(t)]]\quad {\rm and} 
\quad G_i(t)\equiv\nabla_i(|{\bf F}(t)|^2)
\nonumber
\ee
to simplify notations. The time-dependent form of the
four FSI derived in \cite{chin02} are: 
\begin{equation}
{\cal T}_{A}^{(4)}(\epsilon)\equiv 
  {\rm e}^{ {1\over 6}\epsilon V(t+\epsilon)}
  {\rm e}^{ {1\over 2}\epsilon T} 
  {\rm e}^{ {2\over 3}\epsilon \widetilde V(t+\epsilon/2)} 
  {\rm e}^{ {1\over 2}\epsilon T} 
  {\rm e}^{ {1\over 6}\epsilon V(t)},
\label{foura}
\end{equation}
with $\widetilde V$ defined by
\be
\widetilde V(t)=V(t)+{1\over 48}\epsilon^2 U(t),
\label{superv}
\ee
corresponding to a modified force
\be
\widetilde {\bf F}(t)={\bf F}(t)+{1\over 48}\epsilon^2{\bf G}(t).
\label{modfa}
\ee
Given ${\bf p}_0={\bf p}(t)$ and ${\bf q_0}={\bf q}(t)$,
transcribing each operator in (\ref{foura}) according to (\ref{maps}) 
yields the following explicit algorithm 4A for advancing the system 
forward to from $t$ to $t+\epsilon$,
\begin{eqnarray}
{\bf p}_1&=&{\bf p}_0+{1\over 6}\epsilon\, {\bf F}({\bf q}_0,t)\nonumber\\
{\bf q}_1&=&{\bf q}_0+{1\over 2}\epsilon\, {\bf p}_1\nonumber\\ 
{\bf p}_2&=&{\bf p}_1+{2\over 3}\epsilon\, 
            \widetilde {\bf F}({\bf q}_1,t+\epsilon/2)\label{sima}\\
{\bf q}_2&=&{\bf q}_1+{1\over 2}\epsilon\, {\bf p}_2\nonumber\\
{\bf p}_3&=&{\bf p}_2+{1\over 6}\epsilon\, {\bf F}({\bf q}_2,t+\epsilon).\nonumber 
\end{eqnarray}
The last ${\bf p}$ and ${\bf q}$ are the updated variables, {\it i.e.}, 
${\bf q}(t+\epsilon)={\bf q}_2$ and ${\bf p}(t+\epsilon)={\bf p}_3$.
For long time integration, when intermediate time results are not needed,
the first and last force evaluation are the same and only need to be
evaluated once. Thus algorithm 4A only
requires two evalutations of the force and one evaluation of the force gradient to
be fourth order. Note that forces are to be evaluated at an intermediate time
equals to the sum of time steps of all the preceding $T$ operators \citep{suz93,chin02}. 

Similarly algorithm 4B in operator form is 
\begin{equation}
{\cal T}_{B}^{(4)}(\epsilon)\equiv 
  {\rm e}^{t_2\epsilon T}
  {\rm e}^{{1\over 2}\epsilon  \bar V(t+a_2\epsilon) } 
  {\rm e}^{t_1\epsilon   T}
  {\rm e}^{{1\over 2}\epsilon  \bar V(t+a_1\epsilon) } 
  {\rm e}^{t_0\epsilon T},
\label{fourb}
\end{equation}
where
\begin{equation}
t_0=t_2={1\over 2}(1-{1\over{\sqrt 3}}),
\quad t_1={1\over{\sqrt 3}},
\quad a_1={1\over 2}(1-{1\over{\sqrt 3}}), 
\quad a_2={1\over 2}(1+{1\over{\sqrt 3}}),
\end{equation}
and with $\bar V$ given by
\begin{equation}
\bar V(t)=V(t)+c_0\epsilon^2\,U(t). 
\label{duperv}
\end{equation}
This is just a modified force with a different coefficient
$c_0={1\over 24}(2-\sqrt 3)$. The transcription to an explicit 
algorithm is similar to the 4A case and will not
be repeated. Algorithm 4B requires two evaluations of the force and
two evalutation of the force gradient. After we have discovered the 
one-parameter family of algorithms discussed below, 
we realize that one can improve 4B by eliminating one evaluation of the
gradient force. This is done by concentrating the commutator term at
the center of algorithm: 
\begin{equation}
{\cal T}_{B^\prime}^{(4)}(\epsilon)\equiv 
  {\rm e}^{t_2\epsilon T}
  {\rm e}^{{1\over 2}\epsilon  V(t+a_2\epsilon) } 
  {\rm e}^{{1\over 2}t_1\epsilon   T}
  {\rm e}^{c_0\epsilon^3U(t+\epsilon/2) } 
  {\rm e}^{{1\over 2}t_1\epsilon   T}
  {\rm e}^{{1\over 2}\epsilon  V(t+a_1\epsilon) } 
  {\rm e}^{t_0\epsilon T}.
\label{fourbp}
\end{equation}
Just to be sure that there is no confusions, we transcribe this 
algorithm 4B$^\prime$ as follows
\vskip 2in
\begin{eqnarray}
{\bf q}_1&=&{\bf q}_0+t_0\epsilon\,{\bf p}_0\nonumber\\ 
{\bf p}_1&=&{\bf p}_0+{1\over 2}\epsilon\,{\bf F}({\bf q}_1,t+a_1\epsilon)\nonumber\\
{\bf q}_2&=&{\bf q}_1+{1\over 2}t_1\epsilon\,{\bf p}_1\nonumber\\ 
{\bf p}_2&=&{\bf p}_1+c_0\epsilon^3\,{\bf G}({\bf q}_2,t+\epsilon/2)\label{simbp}\\
{\bf q}_3&=&{\bf q}_2+{1\over 2}t_1\epsilon\,{\bf p}_2\nonumber\\
{\bf p}_3&=&{\bf p}_2+{1\over 2}\epsilon\,{\bf F}({\bf q}_3,t+a_2\epsilon).\nonumber\\ 
{\bf q}_4&=&{\bf q}_3+t_2\epsilon\,{\bf p}_3\nonumber.
\end{eqnarray}
Again, the last two variables are the updated variables, 
${\bf q}(t+\epsilon)={\bf q}_4$ and ${\bf p}(t+\epsilon)={\bf p}_3$.
In contrast to algorithm 4A, 4B$^\prime$ only requires two force and one gradient
evaluations for every update and can be used to produce continuous outputs.  

One of us has shown \citep{chin97} that Ruth's orginal third order 
algorithm \citep{ruth83} simply corresponds to the following
asymmetric third order decomposition:

\begin{equation}
{\rm e}^{\epsilon(T+V)}= 
  {\rm e}^{\epsilon {1\over 3} T} 
  {\rm e}^{\epsilon {3\over 4} V}
  {\rm e}^{\epsilon {2\over 3} T} 
  {\rm e}^{\epsilon {1\over 4} V^\prime},
\label{ruths}
\end{equation}
with $V^\prime=V+{1\over 12}\epsilon^2[V,[T,V]]$. The power of the
operator approach is that one immediately realizes that the third
order error in (\ref{ruths}) can be eliminated by symmetrization, 
since a symmetric decomposition can only have errors of even order in $\epsilon$.
To symmetrize, take two third order algorithms (\ref{ruths}) at half the time
step, flip one over and concatenate.  
Algorithms 4C and 4D correspond to the two possible ways of concatenation:
\begin{equation}
{\cal T}_{C}^{(4)}(\epsilon)\equiv 
  {\rm e}^{ {1\over 6}\epsilon T} 
  {\rm e}^{ {3\over 8}\epsilon V(t+5\epsilon/6)}
  {\rm e}^{ {1\over 3}\epsilon T} 
  {\rm e}^{ {1\over 4}\epsilon\widetilde V(t+\epsilon/2)}
  {\rm e}^{ {1\over 3}\epsilon T} 
  {\rm e}^{ {3\over 8}\epsilon V(t+\epsilon/6)}
  {\rm e}^{ {1\over 6}\epsilon T},
\label{chinc}
\end{equation}
\begin{equation}
{\cal T}_{D}^{(4)}(\epsilon)\equiv 
  {\rm e}^{{1\over 8}\epsilon \widetilde V(t+\epsilon)}
  {\rm e}^{{1\over 3}\epsilon T} 
  {\rm e}^{{3\over 8}\epsilon V(t+2\epsilon/3)}
  {\rm e}^{{1\over 3}\epsilon T}
  {\rm e}^{{3\over 8}\epsilon V(t+\epsilon/3)}
  {\rm e}^{{1\over 3}\epsilon T} 
  {\rm e}^{{1\over 8}\epsilon \widetilde V(t)},
\label{chind}
\end{equation}
where $\widetilde V(t)$ is the same modified force (\ref{superv}) 
used in algorithm 4A. 

Algorithms 4A, 4C and 4B$^\prime$ are special cases of a one-parameter
family of algorithms
\begin{equation}
{\cal T}_{ACB^\prime}^{(4)}(\epsilon)\equiv 
  {\rm e}^{ t_3\epsilon T} 
  {\rm e}^{ v_3\epsilon V(t+a_3\epsilon)}
  {\rm e}^{ t_2\epsilon T} 
  {\rm e}^{ \epsilon W(t+a_2\epsilon)}
  {\rm e}^{ t_1\epsilon T} 
  {\rm e}^{ v_1\epsilon V(t+a_1\epsilon)}
  {\rm e}^{ t_0\epsilon T}
\label{algac}
\end{equation}
parametrized by $t_0$. Given $t_0$, the rest of the 
coefficients are:
\begin{equation}
t_1=t_2={1\over 2}-t_0,\quad
t_3=t_0,\quad 
v_1=v_3={1\over 6}{1\over{(1-2 t_0)^2}},\quad 
v_2=1-(v_1+v_3),\quad
\label{cofac}
\end{equation}
$a_1=t_0$, $a_2=1/2$, $a_3=1-t_0$, 
with $W(t)$ defined by
\begin{equation}
W(t)
=v_2V(t)+u_0\epsilon^2\,U(t),
\label{vtac}
\end{equation}
and
\begin{equation}
u_0={1\over 12}\biggl[1-{1\over{1-2t_0}}+{1\over{6(1-2t_0)^3}}\biggr].
\label{acofac}
\end{equation}
All coefficients are positive for $t_0$ in the range
$[0,{1\over 2}(1-{1\over{\sqrt 3}})\approx 0.21]$.
At $t_0=0$, one has algorithm 4A. 
At $t_0=1/6\approx 0.167$, one has algorithm 4C. 
At the upper limit of $t_0= {1\over 2}(1-{1\over{\sqrt 3}})$,
$v_2=0$, and one recovers algoirthm 4B$^\prime$.
One can therefore changes continuously from algorithm 4A to 4C to 4B$^\prime$
by varying the parameter $t_0$. This is very useful in cases where
any two of the above algorithms have convergence errors of opposite signs. 
By varying $t_0$, one can set that error to zero with no additional 
computational effort. For solving the gravitational three-body problem in
the next section, we use $t_0=0.138$, which is intermediate between that
of 4A and 4C. 

There is also a one parameter family of algorithms connecting algorithms 4B,
a variant of 4D and 4A \citep{chin02}. This family generally requires 
four force evaluations plus the force gradient. We will
not consider it further in this work.   

\section{The Circularly Restricted Three-Body Problem}

We compare the efficiency of these FSI by solving the
planar circularly restricted 3-body problem  
defined by
\be 
{{d^2{\bf r}}\over {dt^2} }={\bf a}({\bf r},t)\equiv
-{1\over 2}\Big[
 {\bf a}_1({\bf r},t)
+{\bf a}_2({\bf r},t)
\Big]
\label{three}
\ee
where for $i=1,2$,
$$
{\bf a}_i({\bf r},t)={ {{\bf r}-{\bf r}_i(t)}\over{S_i^3} }
\quad {\rm and} \quad
S_i=|{\bf r}-{\bf r}_i(t)|.
$$
The two attractive centers ${\bf r}_1(t)$ and ${\bf r}_2(t)$ orbit about 
the origin in circles with angular velocity $\omega =1$: 
$${\bf r}_1(t)=-{1\over 2} (\,\cos(t),\sin(t)\,),$$
$${\bf r}_2(t)={1\over 2} (\, \cos(t),\sin(t)\, ).$$
The gradient term is no more complicated than the force itself,
\be
 \nabla(|{\bf a}|^2)
 =-{1\over 2}\Big[C_1{\bf a}_1
 +C_2 {\bf a}_2
 \Big]
\label{grad}
\ee
with 
$$C_1= 2S_1^{-3}+3S_1{\bf a}_1\cdot{\bf a}_2-S_2^{-3}$$
$$C_2= 2S_2^{-3}+3S_2{\bf a}_1\cdot{\bf a}_2-S_1^{-3}$$
We solve the problem directly in the space-fixed frame
in which the force, or the acceleration field as defined above,
is explicitly time-dependent. The initial condition  
${\bf r}_0=(0,0.0580752367)$ and ${\bf v}_0=(0.489765446,0)$
produce an intricate and lengthy ``Chinese Coin" orbit,
useful for testing algorithms. The actual period is $18\pi$, 
but in the present case where the two attracting centers are 
indistinguishable, the orbit repeats after $9\pi$. 
For this work we will consider the orbital period to be $P=9\pi$. 

In Fig. 1 we compare the trajectory after three periods ($t=27\pi$)
using the Forest-Ruth algorithm and the our 4B$^\prime$ 
algorithm. In Figure 2 we compare the same trajectory using algorithm M
and algorithm 4C. M is McLachan's fourth order,
four force-evaluation algorithm \citep{mcl95}:
\be
{\cal T}_{M}^{(4)}(\epsilon)\equiv 
  {\rm e}^{ t_5\epsilon T} 
  {\rm e}^{ v_1\epsilon V(t+a_4\epsilon)}
  {\rm e}^{ t_4\epsilon T} 
  {\rm e}^{ v_2\epsilon V(t+a_3\epsilon)}
  {\rm e}^{ t_3\epsilon T} 
  {\rm e}^{ v_2\epsilon V(t+a_2\epsilon)}
  {\rm e}^{ t_2\epsilon T} 
  {\rm e}^{ v_1\epsilon V(t+a_1\epsilon)}
  {\rm e}^{ t_1\epsilon T},
\label{algm}
\ee
where $v_1=6/11$, $v_2=1/2-v_1$, $a_i=\sum_{j=1}^i t_j$ and 
$$
t_1=t_5={{642+\sqrt{471}}\over{3924}},
\quad t_2=t_4={{121}\over{3924}}(12-\sqrt{471}),
\quad t_3=1-2(t_1+t_2).
$$
Algorithms RF and M are representative fourth order symplectic algorithms with
negative intermediate time steps. Both are outperformed by
4B$^\prime$, which only requires two force and one gradient evaluations.
We have also tested the standard four force-evaluation Runge-Kutta and the
three force-evaluation Runge-Kutta-Nystr\"om algorithm \citep{bat99}. 
At the same step size, both Rung-Kutta type algorithms are unstable and 
cannot produce a bounded trajectory. Algorithm 4A and 4B are comparable
to 4B$^\prime$, with 4D besting even 4C because it uses one more force
evaluation. 

The accuracy of each algorithm can be quantitatively assessed by monitoring
the Jacobi constant. The Jacobi constant is
usually given in the co-rotationg frame in which the two attractive centers
are at rest. However, it is just as easy to transform the Jacobi constant
back to the space-fixed frame where it has a simple interpretation.
For our circularly restricted three-body problem defined by (\ref{three}), the
Jocabi constant $J$ is given by
\be
J={\bf v}^2
-{{1}\over{|{\bf r}-{\bf r}_1(t)|} }
-{{1}\over{|{\bf r}-{\bf r}_2(t)|} }
-2 {\bf r}\times {\bf v},
\label{jcon}
\ee
which is twice the difference between the energy and the angular
momentum. For fourth order algorithms, we expect
$J(t)=J_0+\epsilon^4 J_4(t)$ and the step-size independent
error coefficient $J_4(t)$
can be extracted from
$$
J_4(t)=\lim_{\epsilon\rightarrow 0}{{(J-J_0)}\over{\epsilon^4}}.
$$
As one decreases the step size, $J_4(t)$ converges to a fixed shape 
characteristic of the algorithm. In Fig. 3 we show this coefficient function 
for the regular fourth order Runge-Kutta algorithm at a step size of 
$\epsilon=P/50000\approx 0.00056$. In the course of
one period, the third body has five close encounters with the two attractive
centers, resulting in five error spikes at t/P=1/10, 3/10, 5/10, 7/10, and 9/10.
Between each close encounter there are also four minor encounters, resulting 
in barely discernable error blips. Fig. 3 demonstrates the distinction between 
symplectic and non-symplectic algorithm. For symplectic algorithms, the error 
recovers back to zero after each close encounter and remain bounded
after each period. For Runge-Kutta type algorithms, each encounter produces 
an irrecoverable error. The accumulating error then grows linearly in 
time without limit. 

Fig. 3 shows the qualitative difference between symplectic and non-symplectic 
algorithms. In Fig. 4, we show the quantitative difference between forward
and non-forward integrators by examining the detail 
shape of the error spike at t/P=1/10. In Fig. 4a, the left figure, we compare 
the error function $J_4(t)$ generated by algorithms FR, M and Cor with that
of 4A. Cor is a processor or corrector algorithm of the form:
\be
{\cal T}_{Cor}^{(4)}(\epsilon)\equiv 
{\rm e}^{C(\epsilon)}
{\cal T}_{2M}^{(2)}(\epsilon)
{\rm e}^{-C(\epsilon)},
\label{cor}
\ee
where the kernel is a second order algorithm (2M) 
\be
{\cal T}_{2M}^{(2)}(\epsilon)\equiv
{\rm e}^{{1\over 2}\epsilon T}
{\rm e}^{\epsilon V^\prime (t+\epsilon/2)}
{\rm e}^{{1\over 2}\epsilon T}
\label{k2m}
\ee
with same modified force $V^\prime$ as defined by (\ref{modforce}).
This second order kernel has been used in most processor or corrector algorithms 
\citep{mar96,mar97,wis96,bla99}. However, a fourth order
corrector with analytical coefficients seems to have been overlooked in 
all of the above references. This corrector is \citep{chin022}
\ba
{\rm e}^{C(\epsilon)}&=&
{\rm e}^{v_2\epsilon V(t+[t_1+t_2]\epsilon)}
{\rm e}^{t_2\epsilon T}
{\rm e}^{v_1\epsilon V(t+t_1\epsilon)} 
{\rm e}^{t_1\epsilon T},\\
{\rm e}^{-C(\epsilon)}&=&
{\rm e}^{-t_1\epsilon T}
{\rm e}^{-v_1\epsilon V(t-t_2\epsilon)}
{\rm e}^{-t_2\epsilon T}
{\rm e}^{-v_2\epsilon V(t)},
\ea
with 
$$
t_1={1\over{2\sqrt{3}}},
\quad 
t_2=-{ 1\over{ 2^{1/3}\sqrt{3}}},
\quad
v_1={1\over{2\sqrt{3}}}-{ 1\over{2^{4/3}\sqrt{3}}},
\quad
v_2=-{ 1\over{2^{4/3}\sqrt{3}}}.
$$
Note that at the end of ${\rm e}^{-C(\epsilon)}$,
time is to be set to $t-(t_1+t_2)\epsilon$. This corrector is unexpectedly short 
because two error terms having identical coefficients can be 
eliminated by a single corrector operator. This is the fastest 
fourth order algorithm when no intermediate results are needed, 
requiring only one evaluation of the force and its gradient. 
When continuous output is needed, as it is here,
it is the slowest algorithm with five force and one gradient
evaluations. Its error function is a factor of two smaller than RF's,
comparable to that of M, but an order of magnitude greater than that of 4A. 
In Fig.4b, the right figure, we compare
all the FSI considered in this work. Algorithm 4B$^\prime$, with one less
force gradient evaluation, has only half the error of 4B. The 
one-parameter family algorithm ACB$^\prime$ with $t_0=0.138$ has only 
one-third the error of 4C. The error height of ACB$^\prime$ is
fully two orders of magnitude smaller than those of non-forward 
symplectic integrators. In table \ref{tab1}, we give the inverse
error height of each algorithm, $1/|h_i|$, normalized to that of RF.
Thus, algorithm ACB$^\prime$'s error height is $\approx 300$ times 
smaller than that of RF and $\approx 150$ times smaller than 
that of algorithm M.      

To further assess the convergence properties of these algorithms, we 
examine the convergence of the third body's energy after one period 
as a function of the step size $\epsilon$. The results are plotted in 
Fig.5. All can be very well fitted by single fourth order monomial 
$c_i\epsilon^4$ as shown by solid lines. One can basically read off 
the efficiency ranking of each algorithm from left to right following 
flatness of the convergence curve. The convergences of RK and RKN 
are similar, despite the fact that RK uses one more force evaluation. 
For this calculation, algorithm Cor is very efficient, requiring 
only one evaluation of the force and its gradient to best FR. 
But surprisingly, the second order kernel algorithm 2M seemed fourth 
order even without the use of correctors! It is nearly 
indistinguishable from the complete corrector algorithm Cor. There is
a plausible explanation for this behavor. The choice of the
modified force (\ref{modforce}) is such that the second order error
terms [T,[V,T]] and [V,[T,V]] have exactly the same coefficient but
opposite in sign.
\be
{\rm e}^{{1\over 2}\epsilon T}
{\rm e}^{\epsilon V^\prime}
{\rm e}^{{1\over 2}\epsilon T}
={\rm e}^{\epsilon\big (\,T+V
     +{1\over 24}\epsilon^2\,[T,[V,T]]
	 -{1\over 24}\epsilon^2\,[V,[T,V]]+\cdots\,\big )}, 
\label{k2mp}
\ee
Evidently, the magnitudes of [T,[V,T]] and [V,[T,V]] are very
nearly equal and their associated errors cancel one another. 
We have explicitly checked that depending on whether the coefficient 
of the modified force is less 
or greater than 1/24, the energy error is
positive or negative respectively,  
in accordance with (\ref{k2mp}).  
 
As in the Jacobi constant case, all forward
symplectic algorithms perform better than non-forward ones.
Interestingly, algorithm 4ACB$^\prime$ is substantially better
than 4A but not 4C. Algorithm 4C and 4D appear to be anomalous; 
their error is definitely fourth order, but with exceedingly small 
coefficients $c_i$. In table 1, we give the
inverse of error coefficient $1/|c_i|$ normalized to that of RF.
Thus the error coefficient of RK or RKN is ten times larger than that of RF,
while the error coefficient of 4C is nearly 2200 times smaller than that of RF.
To give an more quantitative comparison, one should factor in the
number of force and gradient evaluations. Although it is not unreasonable 
to assume that the gradient would require more effort than
evaluting the force, when the force and gradient are 
evaluted at the same time, the additional effort to compute the
gradient, as shown by (\ref{grad}), is minimal. A comparison with RF 
and M, which requires three and four force evaluations respectively,
should give a reasonable gauge of FSI's effectiveness. Taking the
fourth root of $1/|c_i|$ gives the effective step size relative to
RF for attaining the same error. Thus in the case of 4C one can use 
steps size $(2200)^{1/4}\approx 7$ times as large as RF's, 
four times as large as M's and 12 times as large as RF's.  
    
Some of the observed anomalous behavior may be due to the fact that we 
have choosen an end point that is too ``tame". After one period, the third body 
returns to a position that's far from both attractors 
with no substantial error in the Jacobi constant. To test this hypothesis, 
we compute the energy again at mid period, when the third body has a close 
encounter and the error in the Jacobi constant is at a peak. This is shown 
in Fig.6. One immediately sees that in this case the kernel algorithm 2M 
converges quardratically as it should and is distinct from the fourth 
order corrector algorithm Cor. Moreover, the convergence of
both 4C and 4D is now clearly fourth order and bested by the optimized
algorithm 4ACB$^\prime$. However, fourth order fits to 4B$^\prime$ and 
4ACB$^\prime$ are exceedingly unnatural, with very small coefficients. 
Both can be well fitted with a single {\it eighth} order curve as shown.
The inverse of the error coefficient is a again given in Table \ref{tab1}.
Despite some inexplicable behaviors, it is clear from Table \ref{tab1} that
forward symplectic integrator as a class, can be orders of magnitude more
efficient than non-forward integrators.	
 
\section{Conclusions}

Forward symplectic algorithms are the only composition algorithms
possible for solving evolution equations with a time-irreversible 
kernel \citep{for01,auer01}. However, even for reversible equations 
FSI have been shown to be far more efficient than non-forward symplectic
integrators \citep{chin97,chin02}. In this work, we have further 
demonstrated their efficiency in solving classical dynamical problems
with time-dependent forces. In the circularly restricted 
three-body problem, this class of FSI can have errors orders of magnitude
smaller than non-symplectic or non-forward algorithms.
While FSI can be used for any classical calculations, 
such as molecular dynamcis \citep{ome02}, they are particularly
ideal for doing long time, high precision evolution of 
gravitational few-body problems. The force gradient is easily calculable
and no more time consuming than evaluating the force. In comparison to other 
algoirithms, FSI showed excellent convergence behavior even at close 
encounters. The existence of a parametrized 
family of algorithms allows one to optimize the algorithm
for individual applications. This family of FSI should be applied
to more realistic and more complex astrophysical problems.

While FSI have been largely 
overlooked in the development of classical symplectic integrators, they 
are precisely in accordance with McLachlan and Scovel's recommendation  
\citep{mcl962} that ``derivative" symplectic algorithms should be developed. 
This work suggests that a systematic way of deriving higher order ``derivative" 
algorithms is to devise factorization schemes that 
retain higher order commutators.

Process or corrector algorithms can achieve 
fourth order accurracy with only one evalutation of the force and its gradient.
If 4A or 4B$^\prime$ are used as kernels, one should be able to achieve sixth 
order accuaracy with only two evalutions of the force and one evaluation of 
the gradient. The use of fourth order FSI as kernel algorithms would generate 
a new family of sixth order process symplectic algorithms with 
minimal numbers of force evaluation. This work is currently in progress.
If one does not insist on having purely forward time steps, then many higher
order algorithms can be generated on the basis of these fourth order FSI.
See extensive constructions by \citet{ome02}.
 
At this time, despite intense effort, no sixth order FSI has been found. 
At the other hand, no proof has been presented that this cannot be done.
Clearly research in FSI is still in its infancy
and deserves further studies.

\acknowledgments

This work was supported, in part, by the National Science Foundation
grants No. PHY-0100839.

$^*$ Present address: Department of Mathematics, Texas A\&M University,
College Station, TX 77840.


\clearpage


\begin{figure}
\plottwo{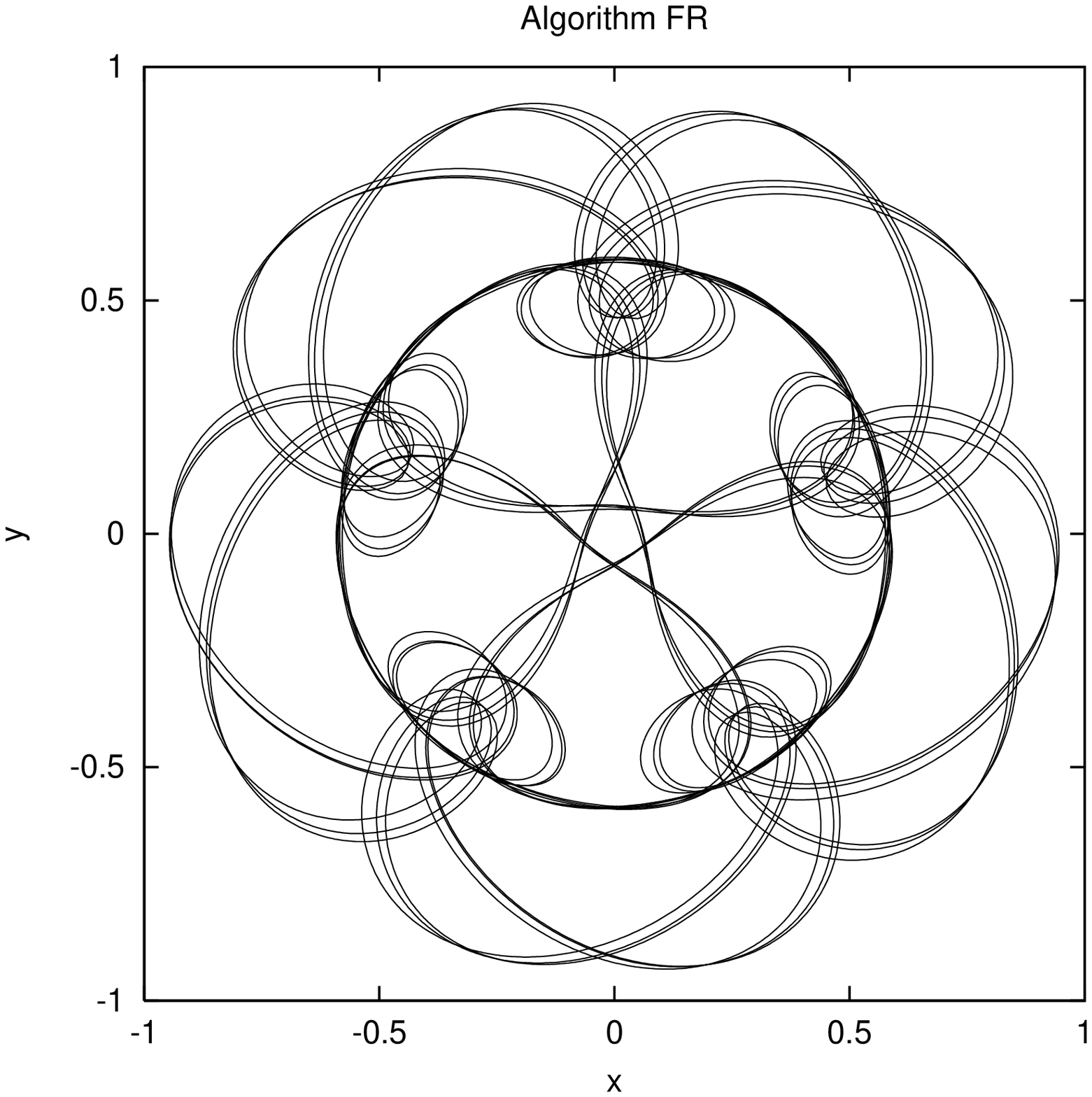}{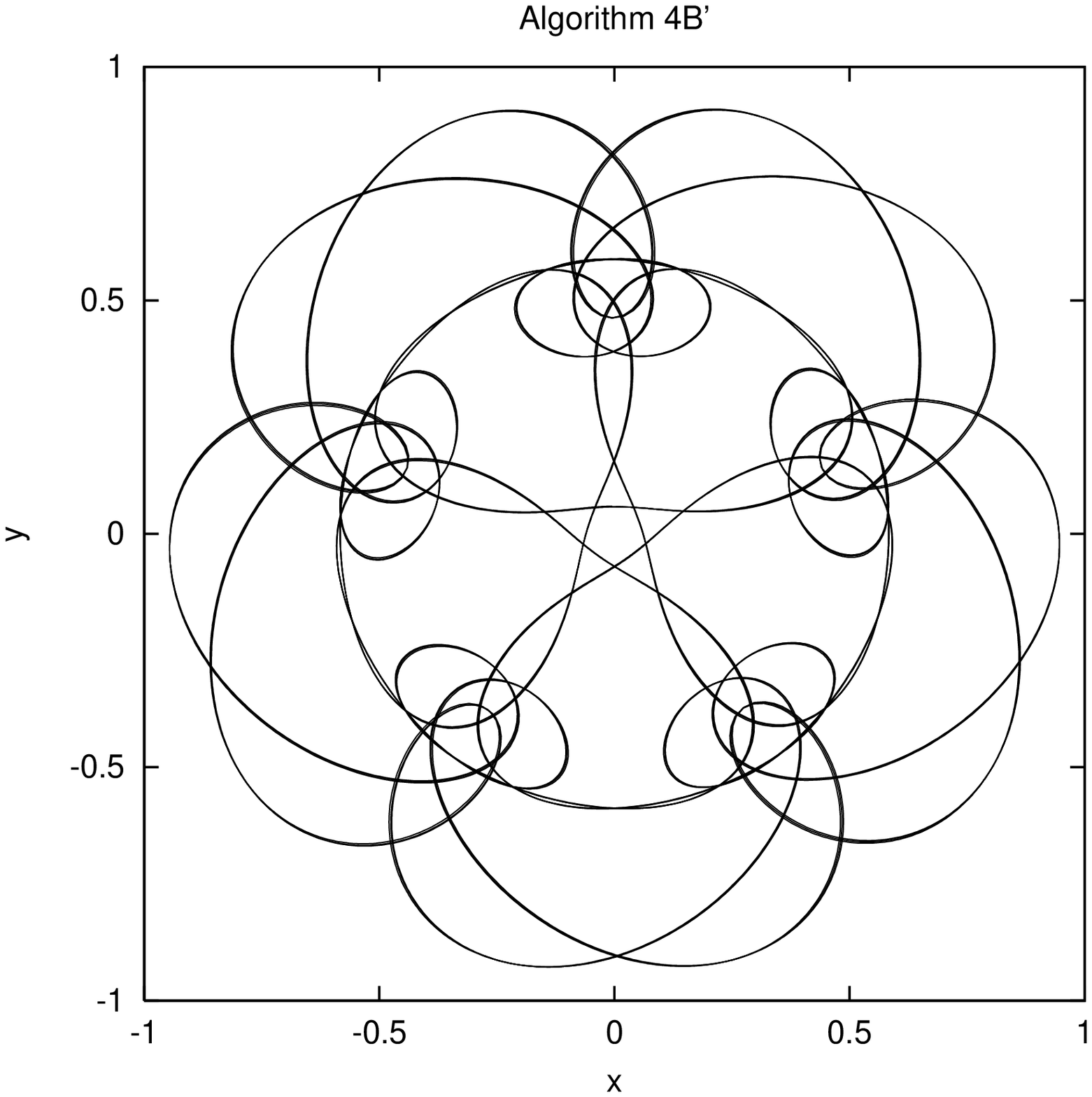}
\caption{The trajectory of the third body in the space-fixed frame
after three orbits in a circularly restricted three-body problem. 
The time step used is very large, $\epsilon=9\pi/5000\approx 0.0057$. 
FR is the Forest-Ruth algorithm which uses three force evaluations per 
update. Algorithm 4B$^\prime$, Eq.(\ref{fourbp}), uses two force and 
one force gradient evaluations. At this large step size, both 
fourth order Runge-Kutta and the Runge-Kutta-Nystr\"om algorithms are 
unstable, producing only chaotic trajectories shooting off to infinity.  
\label{fig1}}
\end{figure}
\begin{figure}
\plottwo{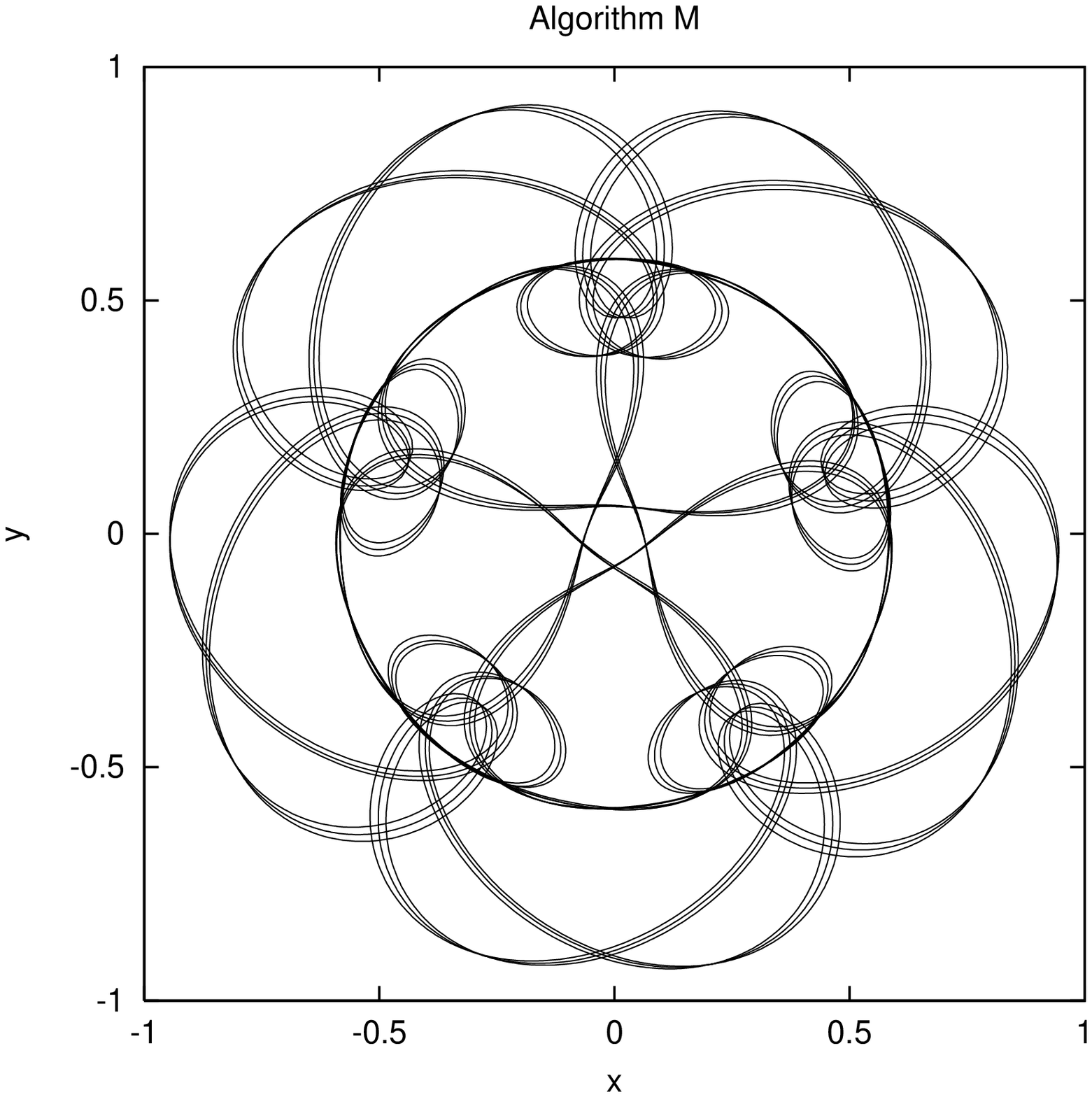}{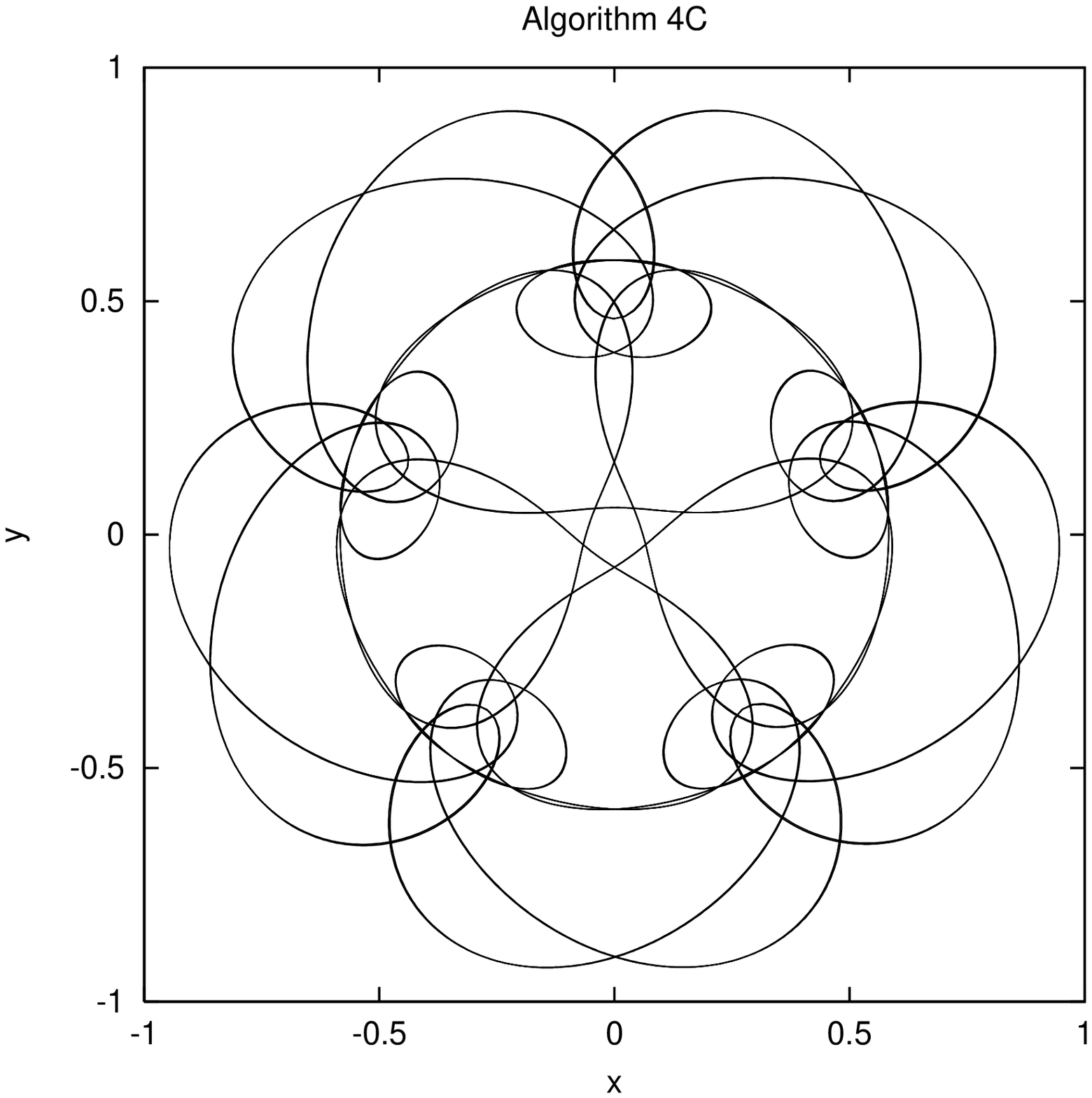}
\caption{Same number of orbit and time step size as in Fig.\ref{fig1}. M is
McLachan's fourth order algorithm which uses four force evaluations per 
update. Algorithm 4C, Eq.(\ref{chinc}), uses three force and one force 
gradient.
\label{fig2}}
\end{figure}
\begin{figure}
\plottwo{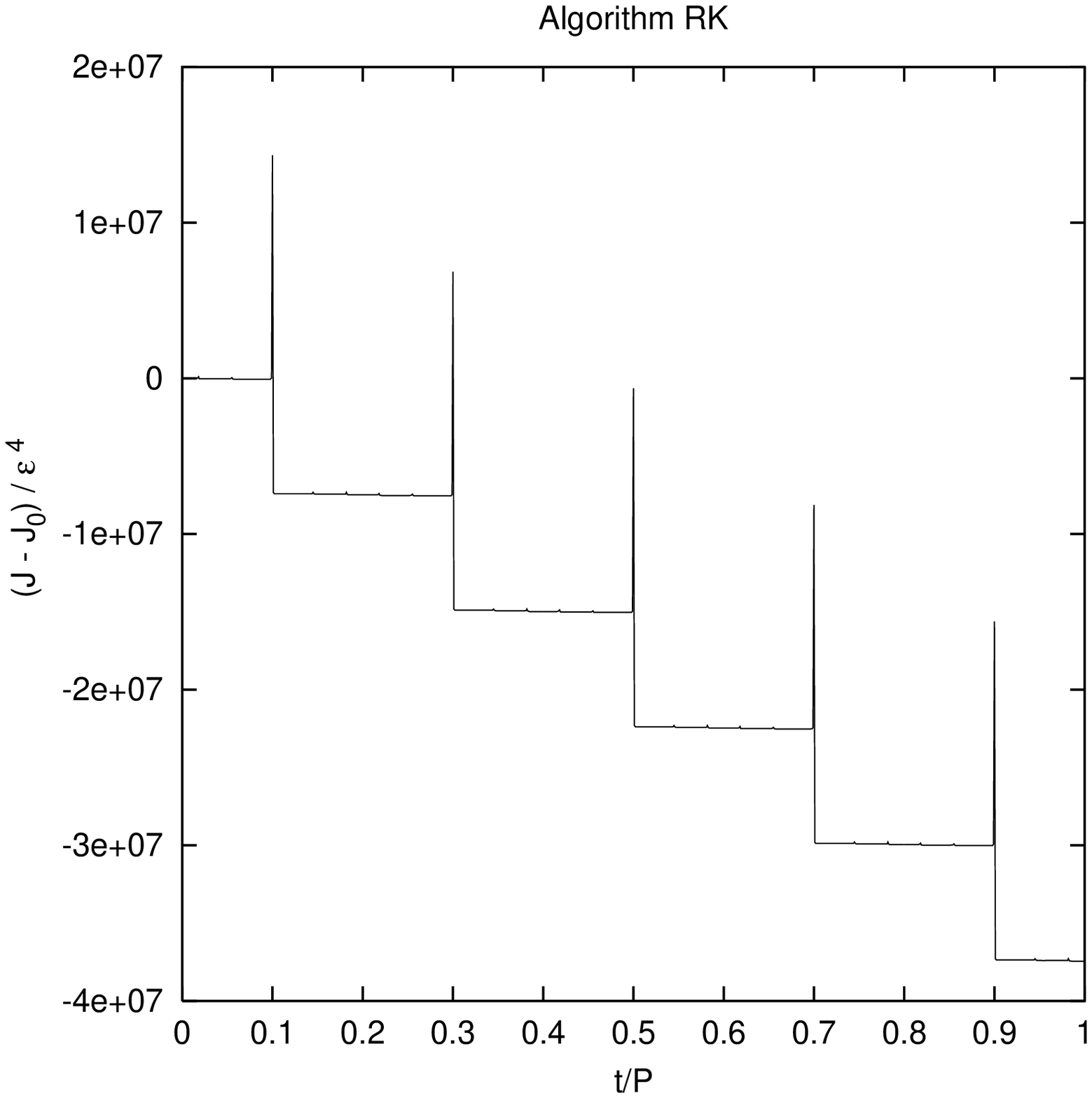}{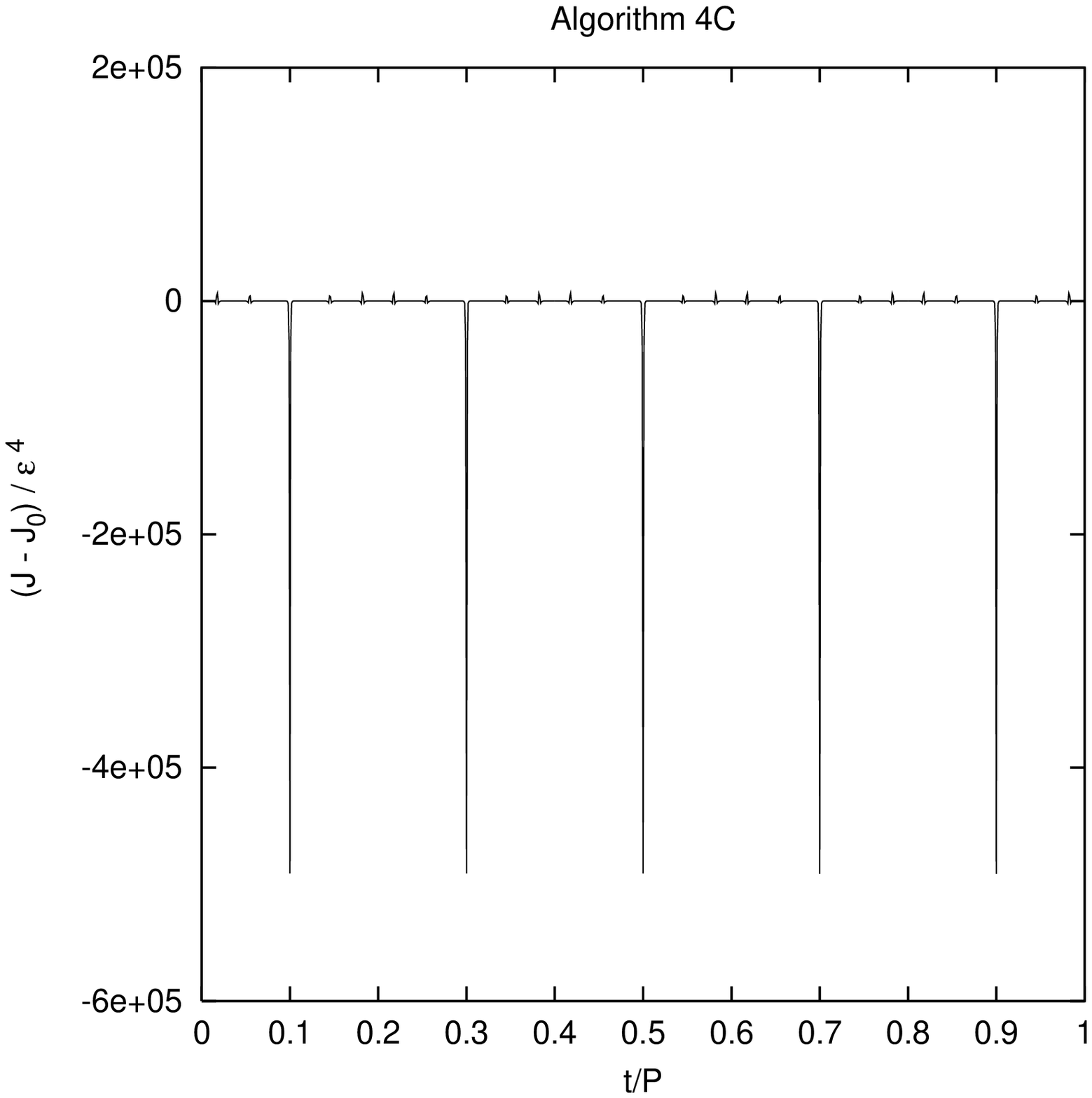}
\caption{The error coefficient of the Jacobi constant as computed by the
fourth order Runge-Kutta algorithm and the forward symplectic algorithm 4C.
Note the relative scale, algorithm 4C's error coefficient is two orders 
of magnitude smaller.
\label{fig3}}
\end{figure}
\begin{figure}
\plottwo{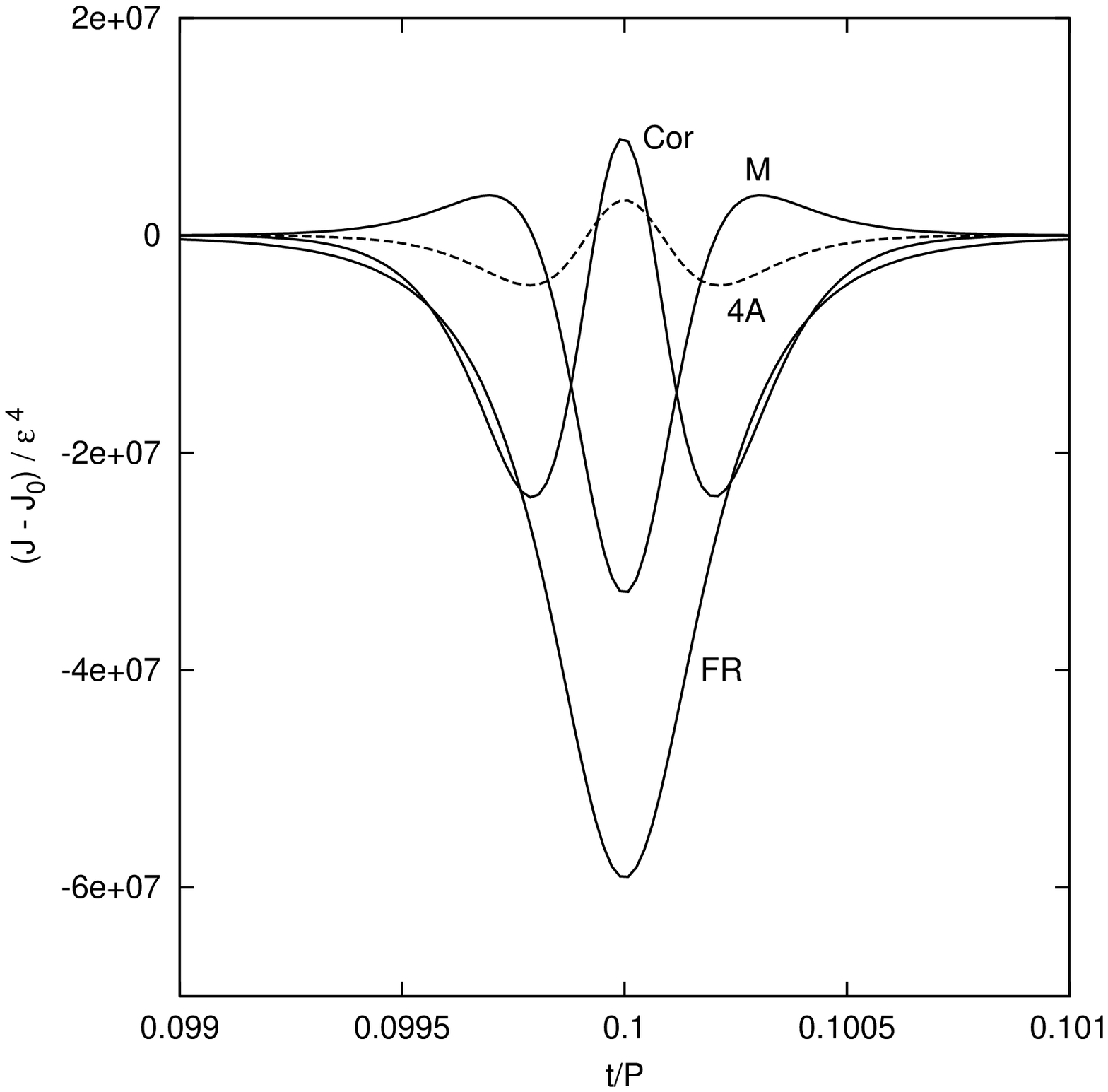}{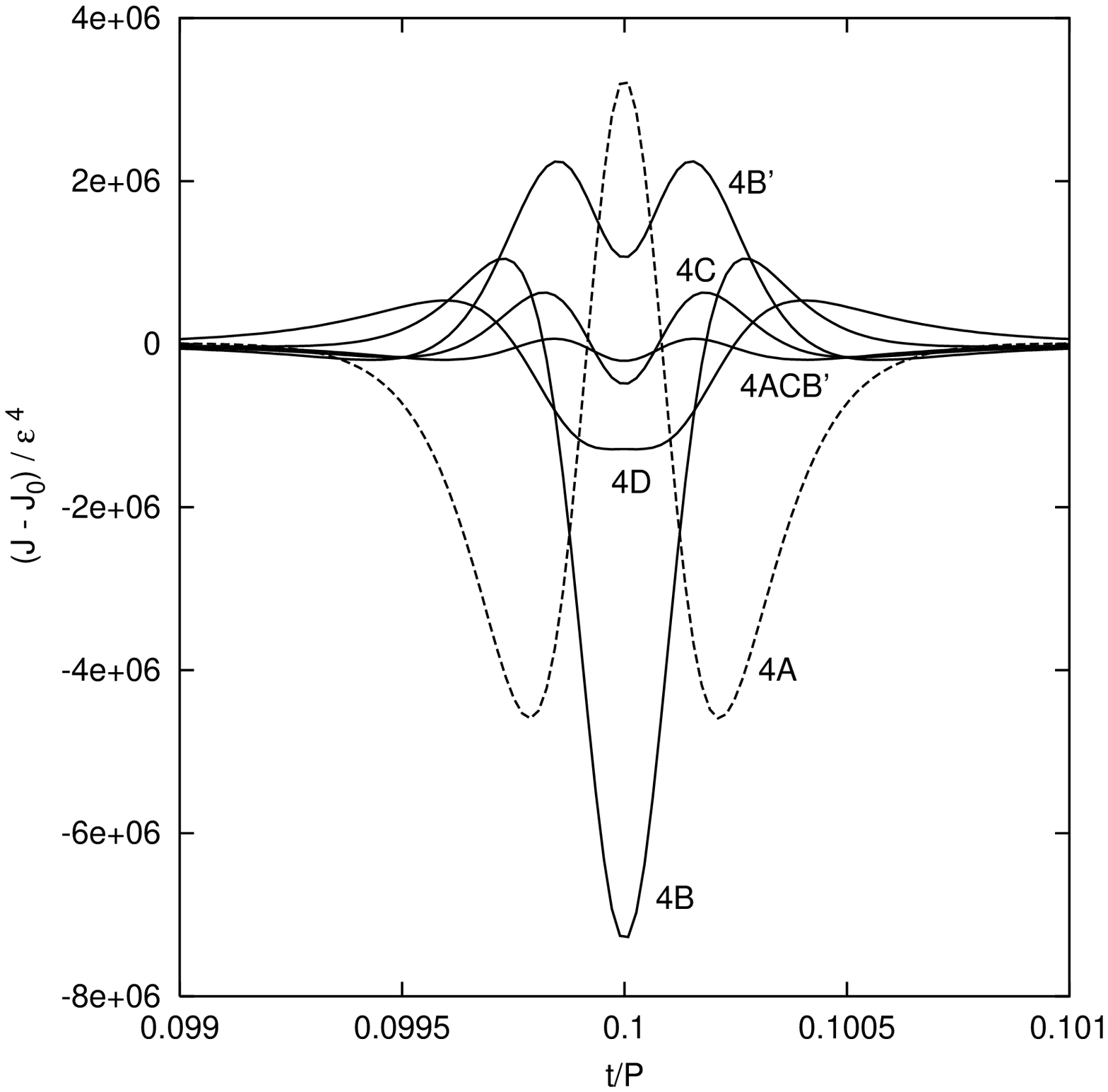}
\caption{The error coefficient of the Jacobi constant at t/p=1/10
due to non-forward symplectic algorithms FR, M and Cor (left figure) 
and forward symplectic algorithms (right figure). Algorithm 4A is drawn with
dotted lines in both for comparison.
\label{fig4}}
\end{figure}

\clearpage

\begin{figure}
\plotone{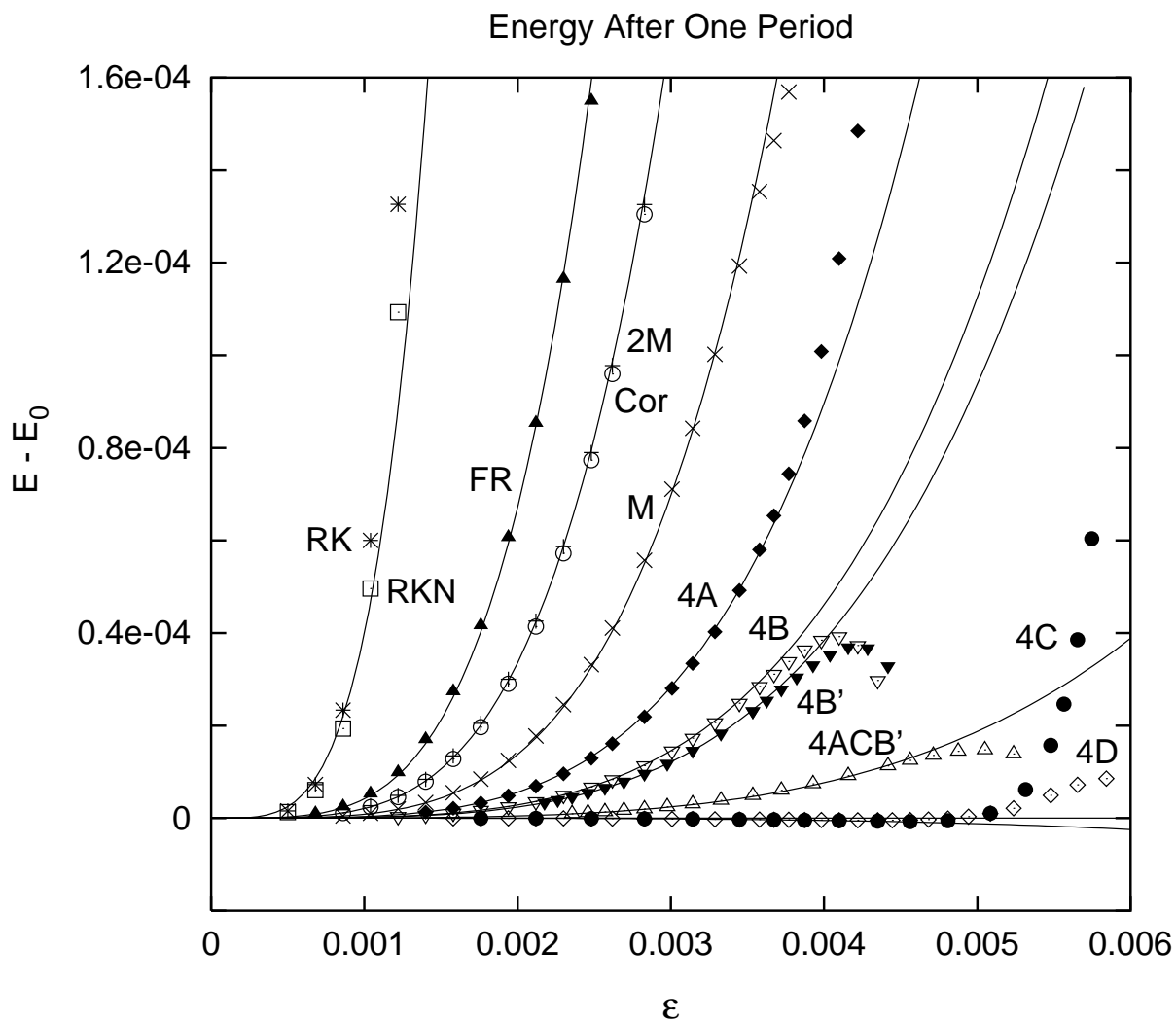}
\caption{The convergence of the energy after one period.
For the ease of comparion, the signs of 4B$^\prime$, RK, and RKN
have been changed from negative to positive. All can be
very well fitted by solid lines of the form $c_i\epsilon^4$.
Though not visible, the convergence curves for 4C and 4D definitely has 
a negative fourth order bend before turning positive. The plus sign 
corresponding to the kernel algorithm 2M, which is nearly indistinguishable 
from the complete corrector algorithm Cor denoted by hollow circles. 
\label{fig5}}
\end{figure}

\clearpage

\begin{figure}
\plotone{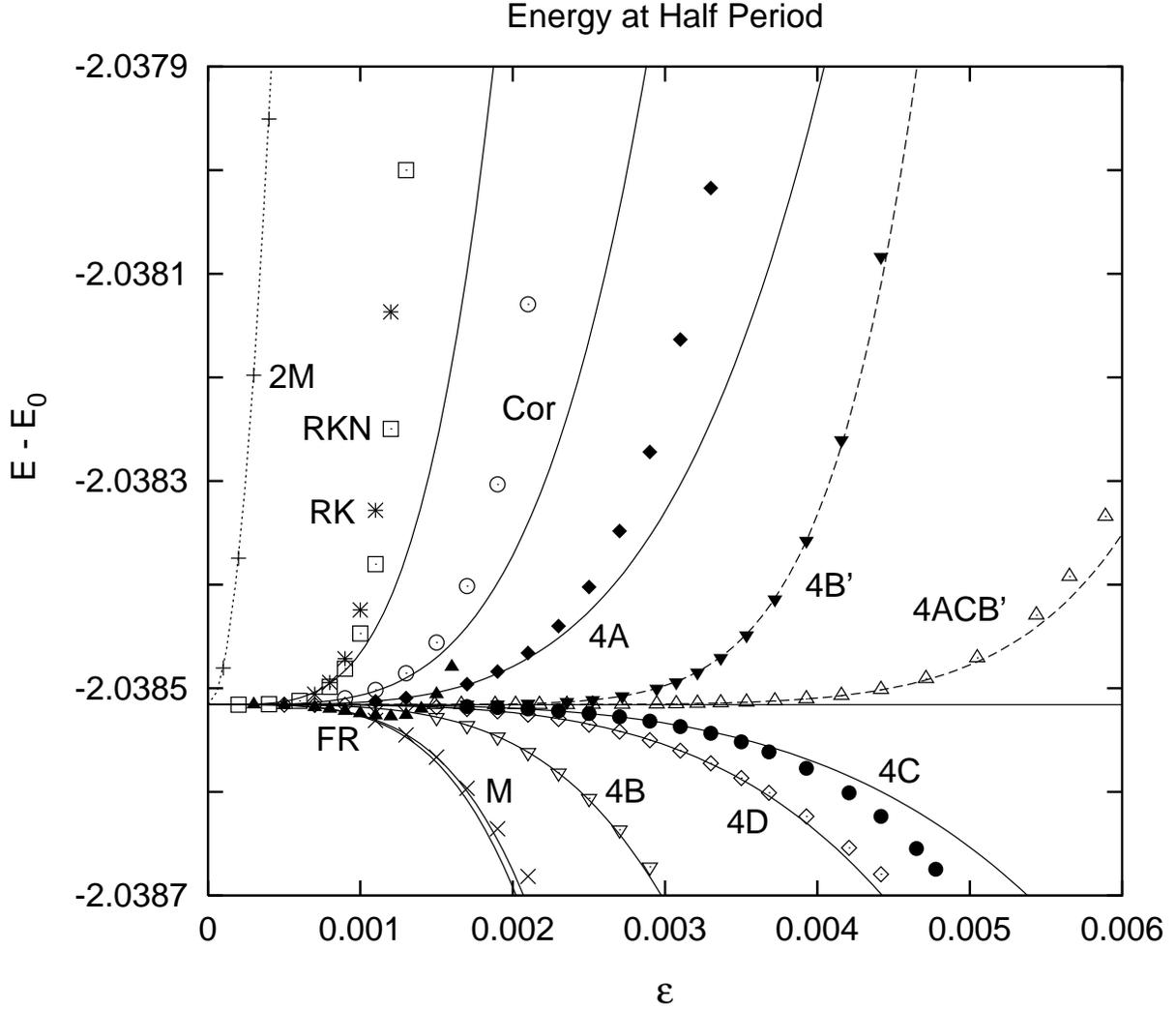}
\caption{The convergence of the energy at mid period where the
error in the Jacobi constant is at its peak. The solid lines are
fourth order fits. The dotted line is a quadratic fit to the kernel 
algorithm 2M. The dashed lines are {\it eighth} order fits to 4B$^\prime$ 
and 4ACB$^\prime$ whose fourth order coefficients are exceedingly small. 
The upright solid triangles are FR's results, which has an initial negative 
fourth order slope, very close to that of M, before turning positive. 
\label{fig6}} 
\end{figure}

\clearpage 

\begin{table}
\caption{The inverse error height in the Jacobi constant 
and the inverse fourth order error coefficient in term of RF's value. 
For example, algorithm
4ACB$^\prime$'s maximum error in the Jacobi constant is 295 times smaller 
than that of FR's and 118 times smaller than that of M, McLachlan's alorithm.
After one period, algorithm 4C's energy error coefficient is 2200 times smaller
than that of FR and 1100 times smaller than that of Cor. At mid period, the fourth
order error coefficient of 4B$^\prime$ and 4ACB$^\prime$ are too small to
be extracted with confidence. Both can be well fitted with an eighth order
error term as shown in Fig.\ref{fig6}. }
\begin{center}
\begin{tabular}{|c|r|r|r|r|r|r|r|r|r|r|}
\tableline\tableline
& RK/RKN & RF & Cor  & M & 4A & 4B & 4B$^\prime$ & 4C & 4D& 4ACB$^\prime$\\
\tableline
$|h_i|^{-1}$ & - & 1 & 2.5  &2 &13 &8  &26 &94 &45 &295 \\	\tableline
$|c_i|^{-1}$ (t=P) & 0.1 & 1 & 2  & 4.9 &12 &23  &28 &2200 &2300 &140 \\
\tableline
$|c_i|^{-1}$ (t=P/2) & 0.3 & 1 & 1.1  & 0.9 &4.3 &4.3  &$\sim$250 &46 &21 &$\sim$2000 \\
\tableline
\end{tabular}
\end{center}
\label{tab1}
\end{table}


\end{document}